\documentclass[twocolumn,pra,unsortedaddress,superscriptaddress,floatfix,citeautoscript,nofootinbib,longbibliography]{revtex4-1}
\usepackage{adjustbox}
\usepackage{dcolumn}% Align table columns on decimal point
\usepackage{latexsym,epsfig,graphicx,dcolumn,subfigure,comment,ulem}
\usepackage{amsmath,eqnarray,amssymb,amsbsy}
\usepackage{xcolor,color}
\usepackage[colorlinks,urlcolor=blue,citecolor=blue]{hyperref}

\begin{document}

\title{Complex Landau levels and related transport properties in 
the strained zigzag graphene nanoribbons}

\author{Zhi-qiang Bao}
\email{zqbao@phy.ecnu.edu.cn}
\affiliation{Key Laboratory of Polar Materials and Devices (MOE), 
	Department of Electronics, East China Normal University,     
	Shanghai 200241, China}
\author{Ju-wen Ding}
\affiliation{Key Laboratory of Polar Materials and Devices (MOE), 
	Department of Electronics, East China Normal University,     
	Shanghai 200241, China}
\author{Junjie Qi}
\email{qijj@baqis.ac.cn}
\affiliation{Beijing Academy of Quantum Information Sciences, 
	West Bld.3, No.10 Xibeiwang East Rd., Haidian District, 
	Beijing 100193, China}

\begin{abstract}

The real magnetic fields (MFs) acting on the graphene can induce flat 
real Landau levels (LLs). As an analogy, strains in graphene can produce  
significant pseudo MFs, triggering the appearance of dispersive pseudo 
LLs. By analyzing the low-energy effective Hamiltonian, we introduce 
the concept of the effective orbital MFs to integrate the real MFs and 
pseudo MFs. Accordingly, we obtain the complex LLs which incorporate the 
real LLs and pseudo LLs, and calculate the related transport properties. 
These concepts enable us to uncover the mechanisms 
driving the fragility of pseudo LLs against disorders and dephasing, 
proving that tuning the real MFs and Fermi energy can effectively 
improve the robust performances. Furthermore, the tunability of the 
valley-polarized currents is also studied, opening up new possibilities 
for the design of valleytronics devices.

\end{abstract}

\maketitle

\section{Introduction}\label{sec:introduction}

Unique in two dimensions, graphene possesses two non-equivalent Dirac points, $K$ 
and $K'$, which leads to the valley degree of freedom ~\cite{grapheneRMP,Beenakker}. 
The two main edge types of graphene nanoribbons (GNRs), which are basically 
one-dimensional structures cut from graphene, are zigzag edges and armchair edges. 
The contrast between the $K$ and $K'$ valleys in the $k-$spaces is one of the primary 
differences between the zigzag GNRs (ZGNRs) and armchair GNRs (AGNRs).  Without 
intervalley scattering, the $K$ and $K'$ valleys are separated and decoupled 
specifically for the ZGNRs in the low-energy limit, making the study of valley 
transport pertinent. However, the $K$ and $K'$ valleys for the AZNRs are both 
projected to the $k$-space $\Gamma$ point, indicating that they are not suitable 
for creating valleytronic devices~\cite{valley1,valley2,pLL1}. As a result, we 
concentrated mostly on the ZGNRs in this work. The ability of ZGNRs to generate 
pseudo-magnetic fields (PMFs) through strains, which in turn leads to the appearance 
of pseudo-Landau levels (PLLs), is another remarkable property of the 
material~\cite{pLL1,pLL2,pLL3,pLL4,pLL5,pLL6,pLL7}. This phenomenon has been identified 
in several noteworthy investigations~\cite{exp1,exp2,exp3,exp4}. Moreover, several 
experimental studies have shown that graphene can resist nondestructively reversible 
deformations up to high values of 
$25\%$-$27\%$~\cite{strength1,strength2,strength3,strength4}, implying that it could 
be a promising material for building novel straintronic devices with the exceptional 
features associated with PMFs.

\begin{figure}[ht!]
\centering
\includegraphics[width=1.0\linewidth]{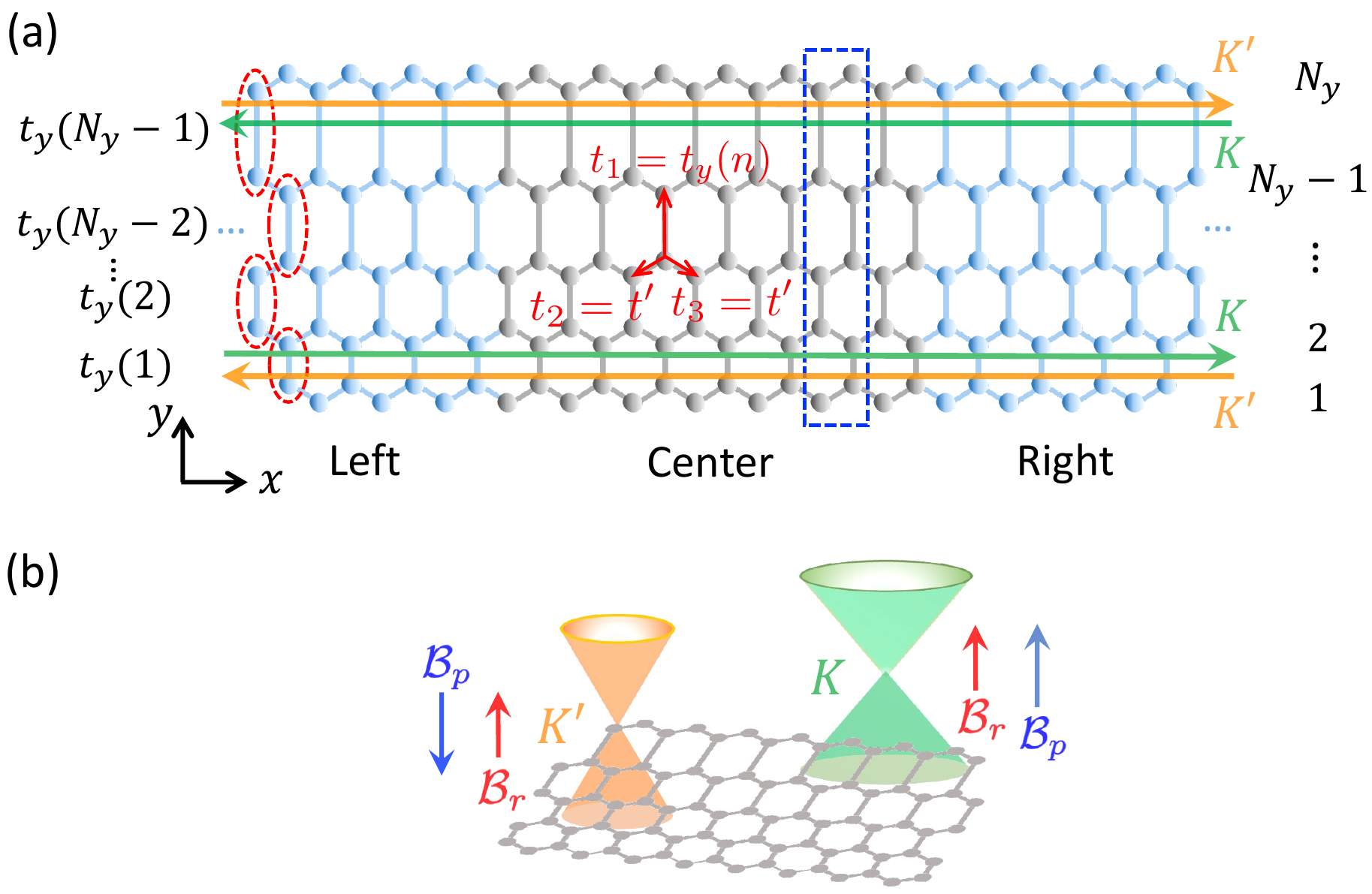}
\caption{(a) Schematic diagram of a two-terminal ZGNR with MIS. The grey (blue) colors represent the central region (leads).
The size of the central region is set as $N_y$ and $N_x$. The blue dotted rectangle represents a primitive cell, which contains $2N_y$ atoms. $N_x$ is the 
number of primitive cells in the $x$ direction.
The green (orange) arrows depict the current for $K$ ($K'$) valley.   (b) illustrates  
the $K$ and $K'$ valleys of the strained graphene with RMFs. }
\label{fig:1}
\end{figure}

A two-terminal ZGNR with the uniaxial strain is  shown in Fig.~\ref{fig:1}(a).
The uniaxial strain of the ZGNR is extended along the $y$ direction.
Accordingly, the hopping coefficients along the $y$ direction $t_y(n)$ are assumed to be a linear 
function of $n$ ($n=1,2,...,N_y-1$). Meanwhile, the hopping coefficients along the $x$ direction 
are considered to be constant~\cite{pLL1,Franz,TB2,HJiang}. As a result, this strain pattern is 
referred to as the monotonic increasing strain (MIS) ~\cite{Bao}, which 
leads to the emergence of a uniform perpendicular PMF.   
Valley-polarized
currents for the $K$ and $K'$ valleys are theoretically predicted in the 
ZGNRs under the influence of the PMFs~\cite{Franz}. Fig.~\ref{fig:1}(b) 
shows the $K$ and $K'$ valleys of the strained GNRs in the real magnetic 
fields (RMFs). As discussed in Sec.~\ref{sec:disorders}, the joint effects 
of the PMFs and RMFs cause the $K'$ ($K$) valley to sink (raise) and get 
narrower (wider). Both the RMFs and PMFs can produce LLs, 
however, the former results in flat LLs while the latter results in 
dispersive ones. Additionally, it is shown by the transport characteristics 
research that the states associated with PLLs and RLLs have distinct 
robust responses to Anderson disorders and dephasing effects. We offer the 
idea of the effective orbital magnetic fields (EOMFs), which result in the 
creation of complex LLs (CLLs), to combine the impacts of RMFs and 
PMFs to explain the transport characteristics of the strained ZGNRs. 
We propose several mechanisms of the intervally and intravalley 
to explain the distinct robust performances for the RLLs and PLLs, and 
 point out that the valley polarization governed by the EOMFs $|\mathcal{B}^{\pm}_{\textrm{eff}}|$ results in the distinct conductance features that are related to the $K$ and $K'$ valleys, respectively.

The paper is organized as follows. We introduce the model and numerical 
methods employed in this work in Sec.~\ref{sec:model}. In 
Sec.~\ref{sec:theory}, we talk about the low energy effective theory and 
introduce the concept of the EOMFs and CLLs. Sec.~\ref{sec:results} 
presents the key findings of our calculations and the corresponding remarks. 
More specifically, in Sec.~\ref{sec:disorders}, we study the effect of the 
Anderson disorders and reveal the mechanisms driving the fragility of PLLs
against disorders. Dephasing effect, which is covered in 
Sec.~\ref{sec:dephasing}, is another barrier to the robustness of the 
CLLs. Sec.~\ref{sec:tunability} discussed the tunability of the valley-polarized 
 currents in the strained ZGNRs. 
Sec.~\ref{sec:conclusion} is the conclusion.

\section{Model and numerical methods}\label{sec:model}

A two-terminal ZGNR with the MIS is illustrated in Fig.~\ref{fig:1}(a). The central region is 
sandwiched between the left ($L$) and right ($R$) leads.  In realistic samples, Anderson disorders and dephasing effects are always 
present. In the following calculations, we suppose that the disorders and 
dephasing effects only exist in the central region. The dephasing effects are easily produced via electron-electron 
interactions, electron-phonon interactions, etc,  and can be tuned by changing the temperature experimentally. Here, we simulate the dephasing effects by 
applying the B\"{u}ttiker's virtual probe assumption\cite{Buttiker}.  The tight-binding Hamiltonian of the ZGNR with MIS in the central region 
can be written as

\begin{align}\label{eq:1}
\mathcal{H}=\sum_{i}\varepsilon_{i}a^{\dagger}_{i}a_{i}-\sum_{\langle ij\rangle}te^{i\phi_{ij}}a^{\dagger}_{i}a_{j},
\end{align}
where $\varepsilon_{i}$ represents the on-site energy, $a^{\dagger}_{i}$ 
and $a_{i}$ represent the creation and annihilation operators, 
and $\left\langle ij\right\rangle $ sums over the nearest neighbors. In the L(R) leads, $\varepsilon_{i}$ equals the Fermi energy $E_F$. In the central region, $\varepsilon_{i}=E_F+W$, where $W$ denotes the disorder strength. Anderson disorders are 
simulated by the onsite energies that are uniformly distributed in 
$[-W/2,W/2]$. If there exists RMFs, the hopping coefficient 
$t$ should have a phase $\phi_{ij}=\int_{i}^{j}\mathbf{A}\cdot d\mathbf{l}/\phi_{0}$ 
with the vector potential $\mathbf{A}$ and the 
flux quantum $\phi_{0}=\hbar/e$. 

 As shown in Fig.~\ref{fig:1}(a), we 
assume that the ZGNRs are only stretched along the $y$-axis, with the 
hopping coefficient $t_y(n)$ being a linear function of $n$. 
 For simplicity, $t_y(n)$ is defined as $t_0$ on the bottom edge and $t_{0}(1-\eta)$ on the 
top edge, respectively. $t_0=-2.75$ eV is the well-known hopping 
coefficient for the normal graphene and $\eta$ is an adjustable variable 
that reflects the strain strength. At any $n$, $t_y(n)$ can be expressed as $t_y(n)=t_{0}(1-\gamma n)$, where $\gamma=\frac{\eta(n-1)}{(N_y-2)n}$. Meanwhile,  $t_2$ and $t_3$
 are  set as $t_0$.  Specifically speaking, previous work 
stated that  
$t_y(n)=t_{0}\exp\left[-\beta\left(\ell_y(n)/a_0-1\right)\right]$~\cite{strength4}, 
where $\ell_y(n)$ is the corresponding bond length along the 
$y$ direction, $a_0=0.142$ nm is the equilibrium bond length of the pristine graphene, and 
$\beta\approx3.37$ is the decay rate. Consequently, $\eta=0.5$ is 
corresponding to the maximum deformation 
strength 20\% and it falls in the regime that is not destructive and 
reversible~\cite{strength1,strength2,strength3,strength4}. 

The conductance is calculated by combining the Landauer-B\"{u}ttiker 
formula with the non-equilibrium Green function method at zero 
temperature~\cite{Landauer,Fisher1981,Meir1992,Jauho1994}. The current in the real or virtual lead can be obtained by 
$I_p=(2e^2/h)\displaystyle{\sum_{q\neq p}} T_{pq}(E_F)(V_p-V_q)$, 
where $p=L,R,1,2,\ldots,N$, $V_p$ is the bias in the lead 
$p$, and $N$ is the number of lattice sites in the central 
region. Here, 
$T_{pq}(E_F)=\textrm{Tr}[\Gamma_p(E_F) G^{r}(E_F)\Gamma_q(E_F) G^{a}(E_F)]$ 
is the transmission function at the Fermi energy $E_F$ from 
lead $q$ to lead $p$, and the line width function is given by 
$\Gamma_{p}(E_F)=i(\Sigma_{p}^{r}(E_F)-\Sigma_{p}^{r \dagger}(E_F))$. 
The retarded Green function is calculated by $G^{r}(E_F)=[G^{a}]^{\dagger}=[E_FI-H-\sum_{p}\Sigma_{p}^{r}(E_F)]^{-1}$, 
where $\Sigma_{p}^{r}(E_F)$ denotes the retarded self-energy 
associated with lead $p$. For the real lead, $\Sigma_{L/R}^{r}(E_F)$ 
can be calculated numerically~\cite{Sancho}; for the virtual lead 
$p$, $\Sigma_{p}^{r}(E_F)=-idp/2$, where $dp$ describes the 
dephasing strength\cite{HJ,YXX}. To drive a current flowing along 
the $x$ direction, a small bias $V=V_{L}-V_{R}$ is added 
between the L and R leads. Once the current $I_{L}$ has 
been obtained, the conductance can be calculated directly by 
$G=(V_{L}-V_{R})/I_{L}$.  The average value of 
$500$ random configurations is used to calculate the conductance.

\section{low-energy effective theory}\label{sec:theory}

For the strained ZGNR, the effective Hamiltonian 
is $\mathcal{H}^{\pm}(\boldsymbol{k})=\boldsymbol{d}\cdot\boldsymbol{\sigma}$~\cite{Franz}, where  

\begin{align}\label{eq:2}
d_{x}^{\pm}=&\hbar v_{F}\left(\pm k_{x}+r_{\mp}\frac{e\mathcal{B}_{p}y}{\hbar}\right), \nonumber \\
d_{y}^{\pm}=&\hbar v_{F}k_{y}\left(r_{\pm}-s_{\pm}\frac{e\mathcal{B}_{p}y}{\hbar}\right).
\end{align}
Here $v_{F}$ is the Fermi velocity of the pristine graphene, $\pm$ represent 
$K$ and $K'$ valleys. $\mathcal{B}_{p}=\frac{\hbar\beta}{2ey}\varepsilon_{yy}$ 
is the PMF induced by the strain, $\varepsilon_{yy}=\partial_{y}u_{y}$ is the strain tensor, and 
$u_{y}$ is the in-plane displacement of carbon atoms along 
the $y$ direction~\cite{Franz,TB2}. 
$r_{\pm}=1\pm\frac{k_{x}}{2}$, 
$s_{\pm}=\left(\frac{3}{2}\pm\frac{k_{x}}{4}\right)$, and we have set 
$a_{0}=1$. It should be pointed out that the Fermi velocities 
are modulated by the strain and should be anisotropic and momentum-dependent. 
By using the same method in Ref.~\cite{HJiang}, we can obtain the Fermi velocities 
of the carriers in the strained graphene 
$v^{s}_{Fx}=\frac{3t_{0}}{2}\sqrt{1+\frac{2\gamma y}{3}-\frac{\gamma^{2}y^{2}}{3}}$ and 
$v^{s}_{Fy}=\frac{3t_{0}}{2}\left(1-\gamma y\right)$.

In the presence of the RMFs, the canonical 
momentum should be replaced by the gauge invariant kinetic momentum, thus 
the $\boldsymbol{d}$ vector changes to 

\begin{align}\label{eq:3}
d_{x}^{\pm}=&v_{F}\left(\pm\Pi_{x}+r_{\mp}e\mathcal{B}_{p}y\right), \nonumber \\
d_{y}^{\pm}=&v_{F}\Pi_{y}\left(r_{\pm}-s_{\pm}\frac{e\mathcal{B}_{p}y}{\hbar}\right),
\end{align}
where $\Pi_{i}=p_{i}+e\mathcal{A}_{ri}$~($i=x,y$). 
Note that $r_{\pm}$ and $s_{\pm}$ 
contain $k_{x}$, thus $p_{x}=\hbar k_{x}$ in $r_{\pm}$ and $s_{\pm}$ 
should also be replaced by $\Pi_{x}$. However, the resulting 
additional terms in $r_{\pm}$ and $s_{\pm}$ can be 
neglected because they are smaller than other relevant terms, 
thus we obtain Eq.~\ref{eq:3}. For a perpendicular RMF, 
we choose the gauge $\mathcal{A}_{rx}=\mathcal{B}_{r}y$ and 
$\mathcal{A}_{ry}=0$, then the Hamiltonian becomes 

\begin{align}\label{eq:4}
\mathcal{H}^{\pm}=v_{F}\left[\sigma_{x}\left(\pm p_{x}\pm e\mathcal{B}^{\pm}_{\textrm{eff}}y\right)+\sigma_{y}p_{y}\left(r_{\pm}-s_{\pm}\frac{e\mathcal{B}_{p}y}{\hbar}\right)\right],
\end{align}
where $\mathcal{B}^{\pm}_{\textrm{eff}}=\mathcal{B}_{r}\pm r_{\mp}\mathcal{B}_{p}$ 
is the EOMF which incorporates the effects of both 
the PMFs and RMFs. Note that $\mathcal{B}^{\pm}_{\textrm{eff}}$ is 
\textit{not} the direct addition of $\mathcal{B}_{r}$ and $\mathcal{B}_{p}$, 
and $r_{\mp}$, the coefficient of $\mathcal{B}_{p}$, is dependent on 
$k_{x}$. This reflects the essential differences between the RMFs and 
PMFs, i.e., RMFs induce flat LLs, but PMFs induce dispersive ones.

By solving the eigenvalue equation

\begin{align}\label{eq:5}
\mathcal{H}^{\pm}\left(\begin{array}{c}
\psi_{A}(y)\\
\psi_{B}(y)
\end{array}\right)=\varepsilon_{\pm}\left(\begin{array}{c}
\psi_{A}(y)\\
\psi_{B}(y)
\end{array}\right),
\end{align}
where $\psi_{A}$ and $\psi_{B}$ are components for the A and B 
sublattices, we can obtain the bulk LLs $\varepsilon_{\pm}$. Using the 
similar method adopted in Ref.~\cite{Franz}, we get

\begin{align}\label{eq:6}
\varepsilon_{\pm}^{2}=2ne\hbar v^2_{F}|\mathcal{B}_{\textrm{eff}}^{\pm}|\left(r_{\pm}+k_{x}s_{\pm}\frac{\mathcal{B}_{p}}{\mathcal{B}_{\textrm{eff}}^{\pm}}\right).
\end{align}

\begin{figure}[ht!]
\centering
\includegraphics[width=1.0\linewidth]{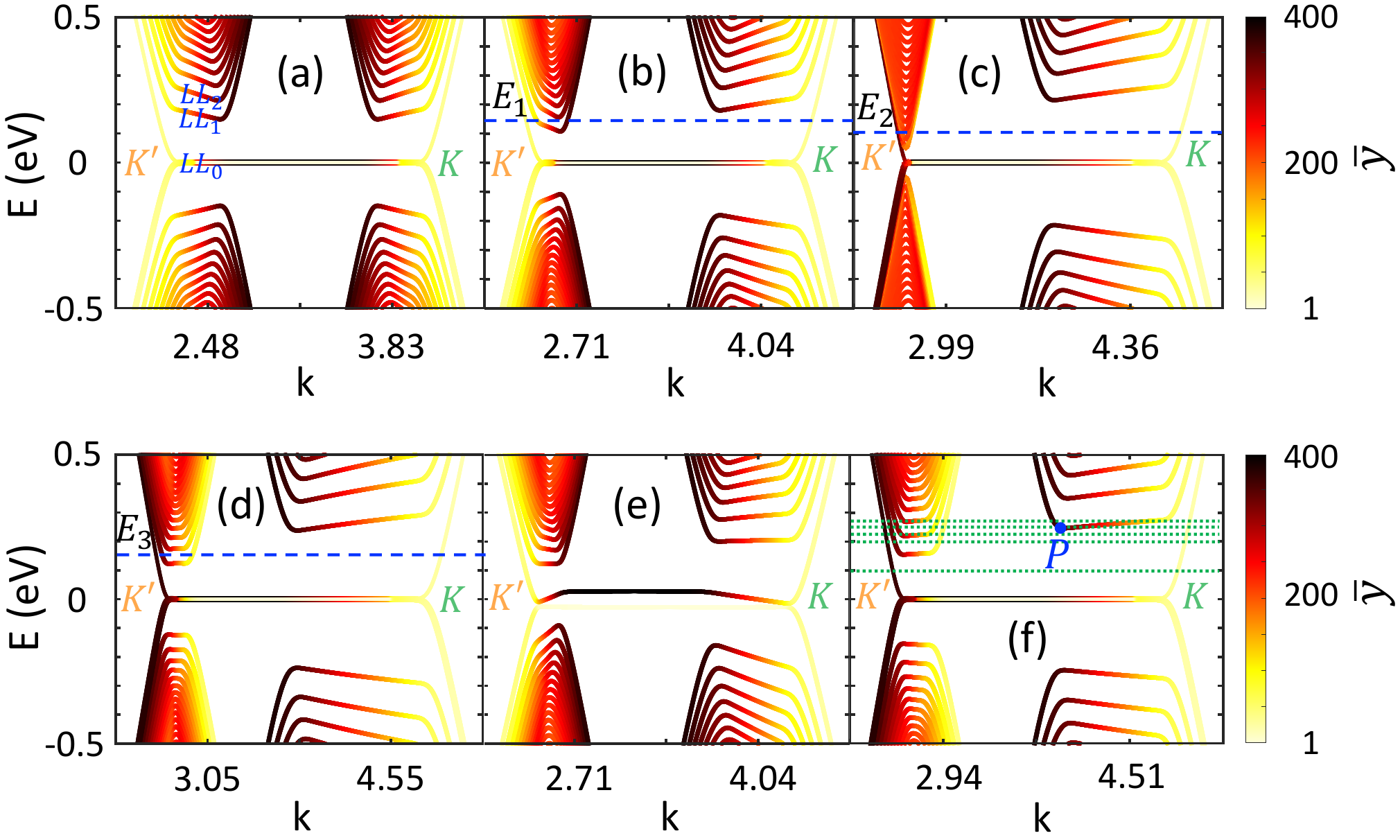}
\caption{(a)-(e) are 
dispersions for the ZGNR with MIS for $\eta=0.5$.  The RMF is 
$\mathcal{B}_{r}=0$ in (a), $\mathcal{B}_{r}=15$~T in (b),  
$\mathcal{B}_{r}=35$~T in (c), and $\mathcal{B}_{r}=50$~T in (d).   
$\mathcal{B}_{r}=15$~T and $\mathcal{E}_{y}=0.02t_{0}$ are adopted in (e).  
$\eta=0.35$ and $\mathcal{B}_{r}=50$~T are adopted in (f), 
in which the green dashed lines labeling several Fermi energies are used to analyze the conductance in Fig.~\ref{fig:3}. The color scale 
represents the expectation value of the $y$ for each eigenstate.  
In all cases, we take $N_y$=200. Specifically, the colors in (a)-(d) and (f) predict 
the degenerate states at $E_F=0$ are localized on both sizes of the sample. The blue dashed lines 
label the Fermi energy $E_F$ in (b)-(d), which are used to 
illustrate the valley currents in Sec. IV C. }
\label{fig:2}
\end{figure}

Because the orientations of the RMF and PMF are opposite for the $K'$ valley, $\mathcal{B}_{\textrm{eff}}^{-}=\mathcal{B}_{r}-r_{+}\mathcal{B}_{p}$ might be 
zero. Thus our solutions for the CLLs are invalid when 
$\mathcal{B}_{\textrm{eff}}^{-}=0$. Actually, 
$\mathcal{B}_{\textrm{eff}}^{-}=0$ is a critical point that the CLLs, 
as well as the edge currents for the $K'$ valley vanish, which also can 
be illustrated in Fig.~\ref{fig:2}(c) and Fig.~\ref{fig:7}(b). 
Furthermore, the result of $\varepsilon^{2}_{-}$ in the $K'$ valley 
is invalid in the vicinity of the singular point 
$\mathcal{B}_{\textrm{eff}}^{-}=0$. For more information, see the 
derivations and discussions in Appendix A.

\begin{figure}[ht!]
\centering
\includegraphics[width=1.0\linewidth]{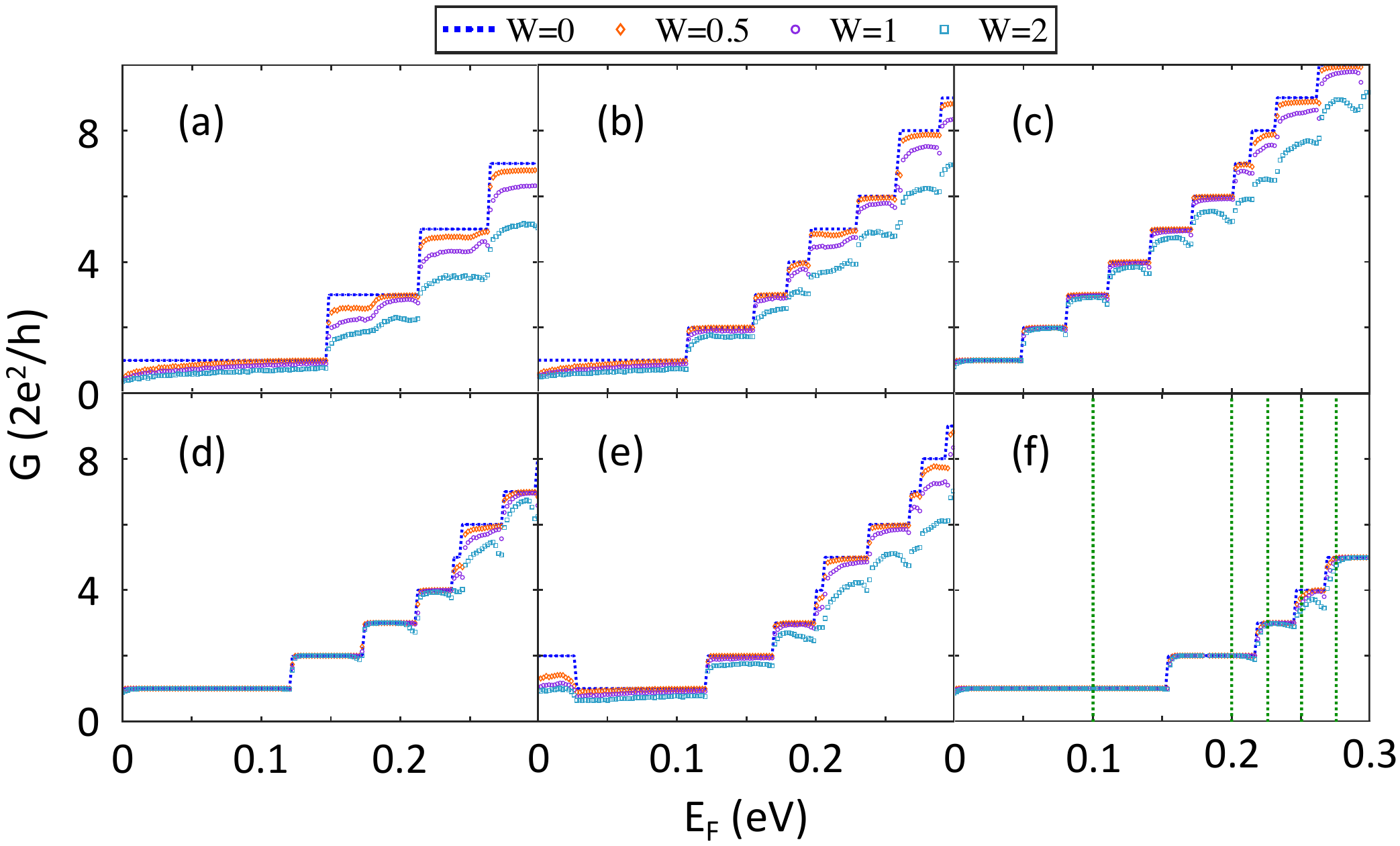}
\caption{(a)-(f): The conductance for the ZGNRs with MIS which are 
one-to-one correspondence with Figs.~\ref{fig:2}(a)-(f). The greed dashed lines in (f) corresponds to those in the energy bands of Fig.~\ref{fig:2}(f). In all cases, we take $N_x$=30 and $N_y$=200.}
\label{fig:3}
\end{figure}

\section{Numerical results and discussions}\label{sec:results}

\subsection{Anderson disorders}\label{sec:disorders}

Next, we examine whether the valley currents in the strained 
ZGNRs are robust against static disorders. 
Figs.~\ref{fig:2}(a)-(e) illustrate the band structures with the strain 
strength $\eta=0.5$. As discussed in Sec.~\ref{sec:theory}, the electrons 
in both valleys encounter the EOMFs $\mathcal{B}^{\pm}_{\textrm{eff}}$. 
When $\mathcal{B}_{r}=0$, $\mathcal{B}^{\pm}_{\textrm{eff}}$ devolves to 
$\pm r_{\mp}\mathcal{B}_{p}$, producing the dispersive and symmetric PLLs in 
Fig.~\ref{fig:2}(a). If $\mathcal{B}_{r}$ is present, the directions of 
the RMF and PMF in the $K$ ($K'$) valley are the same (different). Therefore, 
$|\mathcal{B}^{-}_{\textrm{eff}}|$ keeps decreasing as $\mathcal{B}_{r}$ 
increases before $|\mathcal{B}^{-}_{\textrm{eff}}|$ reaches $0$, and the 
CLLs become lower and narrower in the $K'$ valley as shown in Fig.~\ref{fig:2}(b). 
The CLLs disappear when $|\mathcal{B}^{-}_{\textrm{eff}}|=0$ in 
Fig.~\ref{fig:2}(c). If $\mathcal{B}_{r}$ keeps rising, 
$|\mathcal{B}^{-}_{\textrm{eff}}|$ gradually increases and the CLLs 
reappear in Fig.~\ref{fig:2}(d). Contrary to the $K'$ valley, the CLLs in 
the $K$ valley always become higher and wider because  $|\mathcal{B}^{+}_{\textrm{eff}}|$ 
continues to grow as $\mathcal{B}_{r}$ increase.

\begin{figure}[ht!]
\centering
\includegraphics[width=1.0\linewidth]{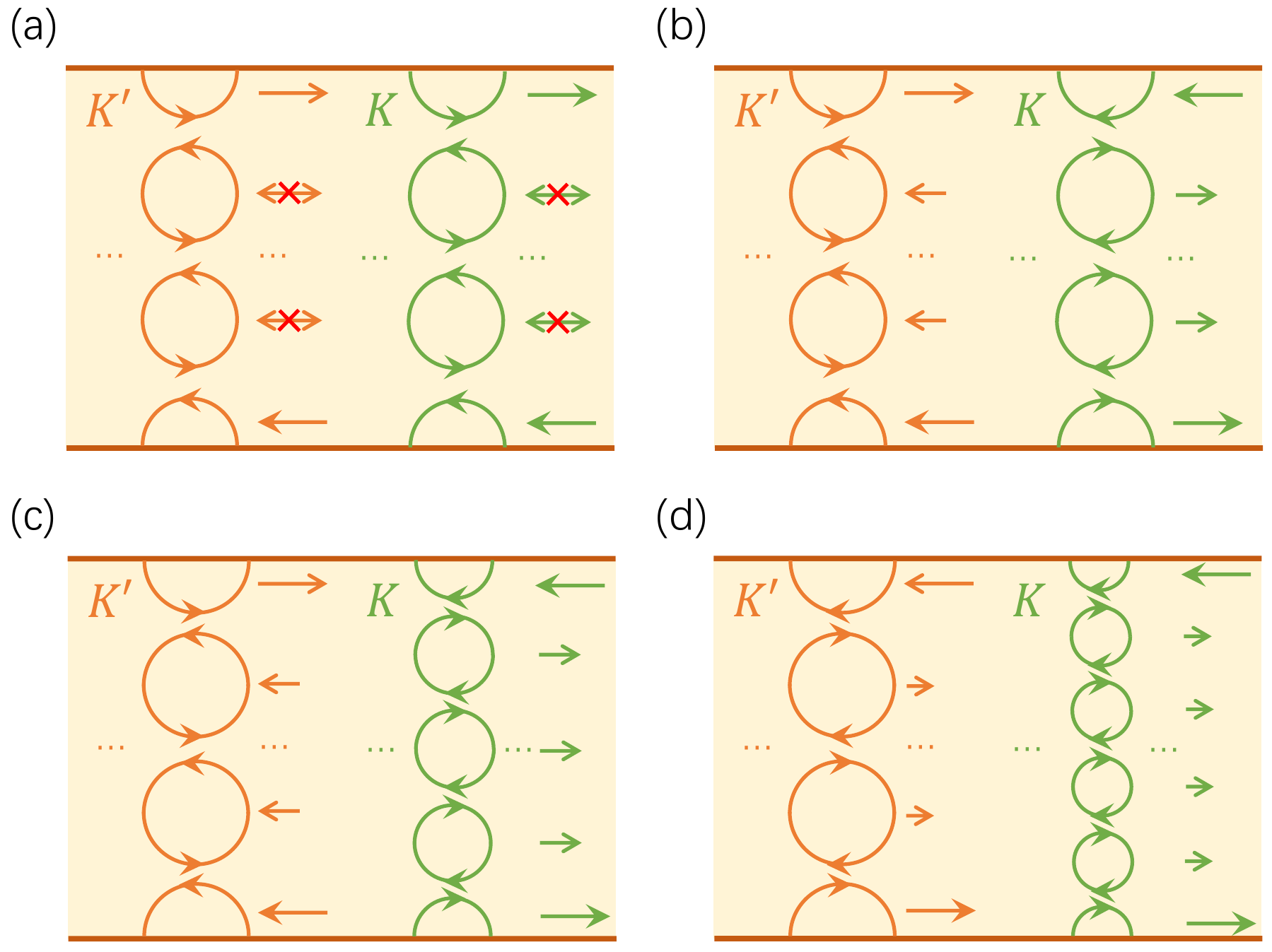}
\caption{The schematic bulk and edge states of (a) the QHE with 
the RMF $\mathcal{B}_{r}$, (b) the QVHE with the PMF 
$\mathcal{B}_{p}$,  (c) the QVHE with both $\mathcal{B}_{r}=15$ T 
and $\mathcal{B}_{p}$ and (d) the QVHE with both $\mathcal{B}_{r}=50$ T  
and $\mathcal{B}_{p}$. The lengths of arrows represent the magnitudes of 
the group velocities, which are the slopes of the energy bands.}
\label{fig:4}
\end{figure}

Fig.~\ref{fig:3}(a) shows that only the first plateau is robust against 
Anderson disorders when $\mathcal{B}_{r}=0$. The similar results have 
been obtained in previous work~\cite{Settnes}, where the authors
attributed the robustness of the first plateau to the polarization of the sublattice.
Figs.~\ref{fig:3}(b)-(d) illustrate the conductance with 
$\mathcal{B}_{r}=15$~T ($\mathcal{B}^{-}_{\textrm{eff}}<0$), 
$\mathcal{B}_{r}=35$~T ($\mathcal{B}^{-}_{\textrm{eff}}\approx0$) and 
$\mathcal{B}_{r}=50$~T ($\mathcal{B}^{-}_{\textrm{eff}}>0$), 
indicating that $\mathcal{B}_{r}$ can improve the robust 
performances because the higher plateaus become more robust. 
\textit{Why are the edges 
states related to the PLLs not robust as that related to RLLs?} There are two main reasons for this. First, the counter-propagating modes 
with spatial overlap and close energies are easily hybridized leading to the enhancement of intervalley scattering. For the quantum Hall effect(QHE) as seen from Fig.~\ref{fig:4}(a), the bulk states 
are localized because the group velocities are zero. 
 The counter-propagating states are entirely separated by the bulk states on the two sides of the sample. Thus,  the conductance plateaus in QHE are robust. According to Fig.~\ref{fig:4}(b), the degenerate counter-propagating states of PLLs overlap in space leading to the enhancement of the intervalley scattering, since the PMFs are opposite between the $K$ and $K'$ valleys due to the time-reversal symmetry. As a result, the conductance plateaus may not be robust due to  hybridizations between the edge-edge 
states, the bulk-bulk states, and the edge-bulk states with spatial overlap.  Additionally, since the counter-propagating modes cannot be spatially separated by the transverse electric field $\mathcal{E}_{y}$, the robustness in Fig.~\ref{fig:3}(e) cannot be considerably improved. It should be noted that the second plateau becomes more robust than it is in Fig.~\ref{fig:3}(a) due to the shift of the valley degeneracy.

Fig.~\ref{fig:4}(c) and ~\ref{fig:4}(d) illustrate the edge and bulk states of quantum valley Hall effect(QVHE) when $\mathcal{B}_{r}$ 
and $\mathcal{B}_{p}$ both exist. In Fig.~\ref{fig:4}(c), $\mathcal{B}_{r}=15$ T 
and $\mathcal{B}^{-}_{\textrm{eff}}<0$, so $|\mathcal{B}^{-}_{\textrm{eff}}|$ ($|\mathcal{B}^{+}_{\textrm{eff}}|$) 
decreases (increases) and the cyclotron radius 
of the electrons in the $K'$ and $K$ valleys gets larger (smaller). Furthermore, The degeneracy of energies between the $K$ and $K'$ valleys has also been lifted, as shown in Fig.~\ref{fig:2}(b). There is only $K'$ valley contributing to the transport, leading to the suppression of  the intervalley scattering, especially for the low-energy regime.
Even for higher energies, the states at the Fermi level are separated in real space when the Fermi energy crosses both two valleys due to the asymmetry between the $K$ and $K'$ valleys. As a result, 
the conductance plateaus become more robust. Fig.~\ref{fig:4}(d) show the case for 
a large $\mathcal{B}_{r}=50$ T and $\mathcal{B}^{-}_{\textrm{eff}}>0$. In this regime, $\mathcal{B}^{-}_{\textrm{eff}}$
makes the electrons in the $K'$ valley counter-rotating. As a result, the directions of valley current for the two valleys coincide, further reducing hybridization and making the conductance plateaus more robust.

\begin{figure}[ht!]
\centering
\includegraphics[width=1.0\linewidth]{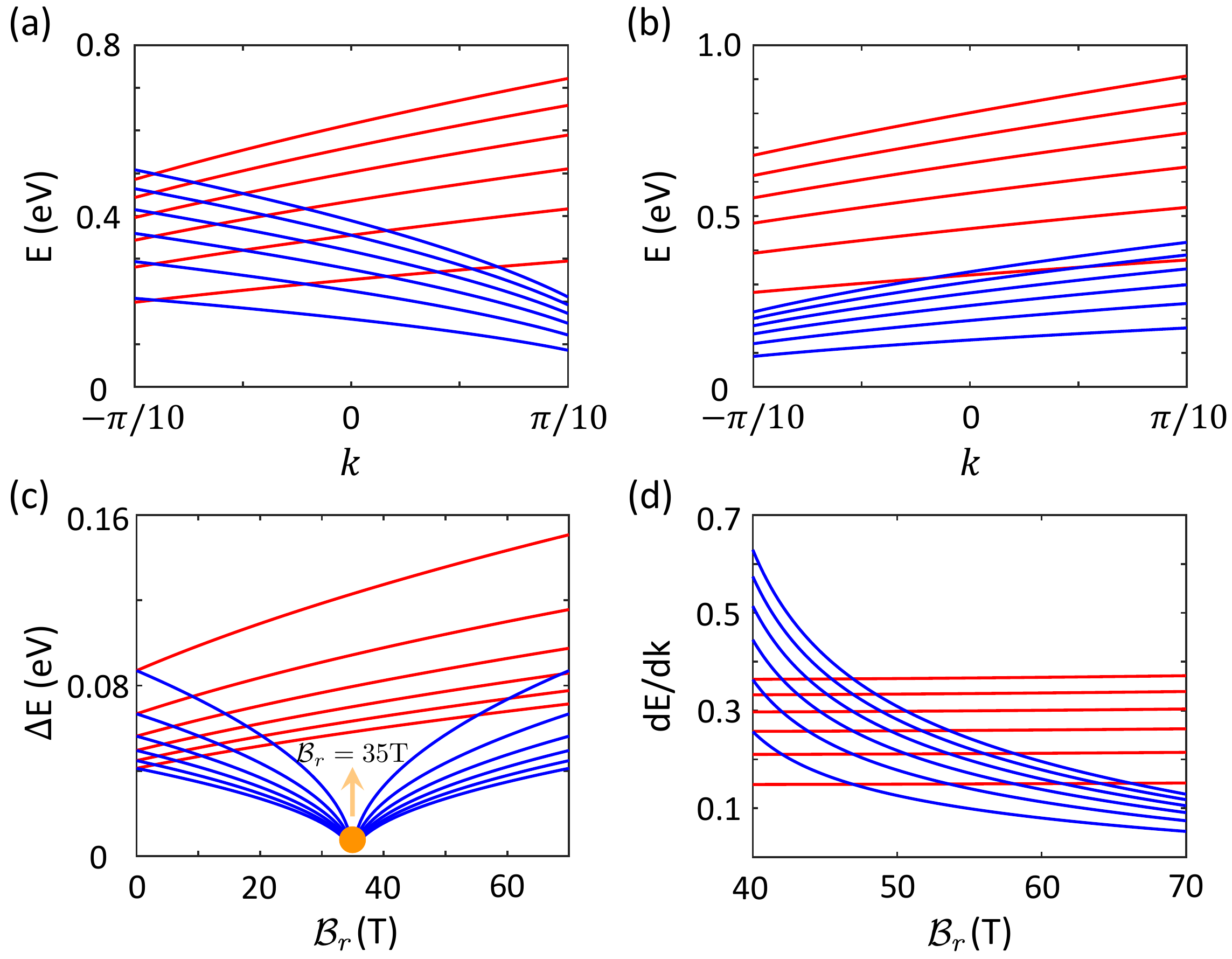}
\caption{(a)-(d) are plotted according to Eq.~\ref{eq:6}. The 
CLLs are shown in the vicinity of the $K$ (red lines) 
and $K'$ (blue lines) valleys with $\mathcal{B}_{r}=15$~T and 
$\mathcal{B}_{r}=50$~T in (a) and (b), respectively. (c) shows the energy 
spacing $\Delta E$ between several adjacent lowest CLLs versus $\mathcal{B}_{r}$ at 
the Dirac point. The small orange circle represents that 
$\mathcal{B}_{r}=35$~T is a singular point. (d) The slopes of the CLLs at the Dirac point.
}
\label{fig:5}
\end{figure}

Second, the intravalley hybridization also has three origins: the bulk-bulk and edge-edge 
hybridizations are weak because they both flow in the same direction; 
the bulk-edge hybridization becomes important because the directions of 
the bulk and edge currents are opposite at least near one edge (see 
Fig.~\ref{fig:4}(b)-(d)). The bulk-edge hybridization can be affected 
by: (i) the degeneracy of CLLs. The results of Figs.~\ref{fig:5}(a) and (b) demonstrate the 
valley polarization in the presence of $\mathcal{B}_{r}$ and are consistent with 
the bands in Figs.~\ref{fig:2}(b) and (d) derived from the tight-binding 
approach. Moreover, we plot the energy spacing of the CLLs at the Dirac 
points ($k_{x}=0$) in Fig.~\ref{fig:5}(c). It can be seen that, in the 
$K'$ valley, $|\mathcal{B}^{-}_{\textrm{eff}}|$ and the energy 
spacing $\Delta E$ monotonically drops and grows as $\mathcal{B}_{r}$ increases when  
$\mathcal{B}_{r}<35$~T and $\mathcal{B}_{r}>35$~T, respectively. 
In the $K$ valley, $|\mathcal{B}^{+}_{\textrm{eff}}|$ always grows as 
$\mathcal{B}_{r}$ increases. Hence, $\Delta E$ between neighboring 
CLLs keeps increasing. If $|\mathcal{B}^{+}_{\textrm{eff}}|$ is larger, the
cyclotron radius in the $K$ valley is smaller, resulting in more bulk states that can be
hybridized with the edge states. (ii) The slopes of the dispersive CLLs, shown in Fig.~\ref{fig:5}(d), characterize the nonzero group velocities that can take part in electronic transport. The counter-propagating modes with 
larger group velocities may lead to stronger scattering. It should be noted that only the $\mathcal{B}_{r}>40$~T
regime is depicted in Fig.~\ref{fig:5}(d) because $\mathcal{B}_{r}=35$~T 
($\mathcal{B}^{-}_{\textrm{eff}}=0$) is a singular point and we address 
the situation in the $\mathcal{B}^{-}_{\textrm{eff}}>0$ regime at this 
time. For the intravalley scattering, 
the bulk and edge states 
related to the dispersive PLLs will inevitably hybridize because it is 
easier for the Fermi energy to cross both the bulk and edge states at 
once. 

Quite interestingly, the seventh plateau is more robust than the sixth 
plateau in Fig.~\ref{fig:3}(d). In order to explore this phenomenon, we 
choose another set of parameters $\eta=0.35$ and $\mathcal{B}_{r}=50$~T in Fig.~\ref{fig:3}(f) 
and obtain similar results that the fifth plateau is more robust than 
the fourth plateau. The green dashed lines plotted in Fig.~\ref{fig:2}(f) are one-to-one correspondence with those in Fig.~\ref{fig:3}(f).
The corresponding relations indicate that the fourth plateau 
is fragile because $E_F$ crosses the first CLL (bulk states) in 
the $K$ valley; however, the higher fifth plateau is still robust 
because only edge states are crossed by $E_F$ in the $K$ valley. 
The hybridization between the counter-propagating modes 
around $P$ (see Fig.~\ref{fig:2}(f)) with close distance in space 
features the fragility of the fourth plateau. The behaviors exhibit 
the polarization of the $K$ and $K'$ valleys, which are determined 
by the EOMFs $|\mathcal{B}^{\pm}_{\textrm{eff}}|$.  In our work, 
the valley polarization has three major aspects: the shift of valley 
degeneracy, the degeneracy of CLLs, and the slopes of CLLs (group 
velocities) in the $K$ and $K'$ valleys. The mechanisms of the 
intervally and intravalley scattering revealed by the preceding 
paragraphs make the different transport behaviors of PLLs and RLLs 
clear. On this basis, valley polarization determined by the EOMFs 
$|\mathcal{B}^{\pm}_{\textrm{eff}}|$ uncovers the different performances 
of conductance which are related to the $K$ and $K'$ valleys, respectively.

Furthermore, it should be pointed out that despite the presence of RMFs in Figs.~\ref{fig:3}(b)-(f), the higher plateaus are still not as robust as the ones in the QHE. On one hand, the CLLs are broadened when the disorder is present. On the other hand, the direct energy spacing $\Delta E$ between two adjacent higher CLLs becomes smaller. Thus, the direct gap between two adjacent CLLs can be smeared out due to the broadening of these states by the disorder. In this regime, bulk-bulk hybridization also plays a role in transport properties. Note that the Rashba spin-orbit coupling (RSOC) and Zeeman energy are not included throughout the work because it has no impact on 
our understanding of physics (See Appendix B for more 
information).

\begin{figure}[ht!]
\centering
\includegraphics[width=1.0\linewidth]{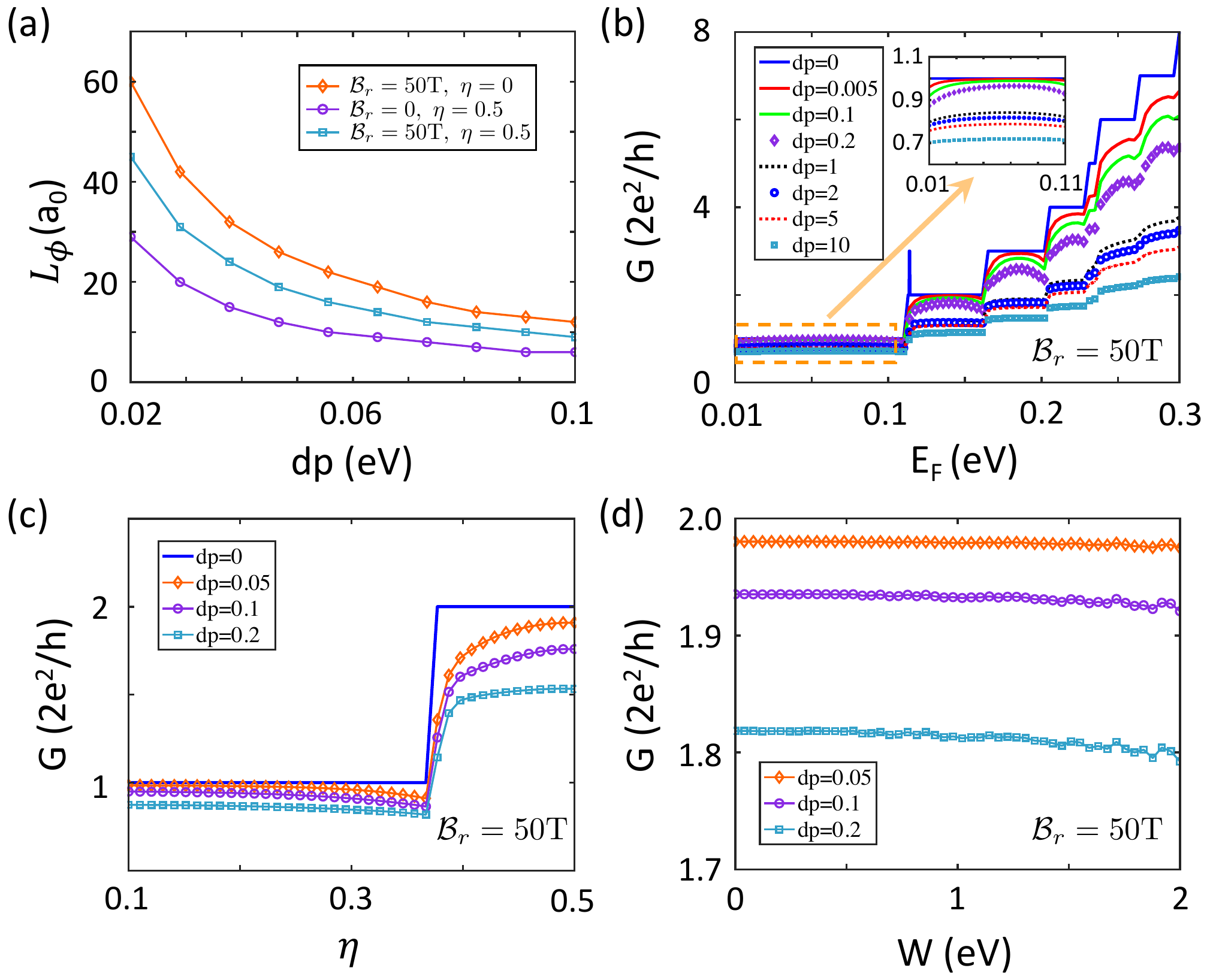}
\caption{(a) The coherent length $L_{\phi}$ vs the dephasing 
strength $dp$ for various EOMFs with $E_F=0.2$ eV.  The conductance $G$ vs (b) the Fermi energy 
$E_F$, (c) the strain strength $\eta$, and (d) the disorder strength $W$ for 
various $dp$, respectively. We set $\eta=0.5$ 
in (b) and (d), $E_F=0.15$~eV in (c) and (d), and $\mathcal{B}_{r}=50$~T 
in (b)-(d). $N_y=200$ in all case; We take $N_x=30$ in (c) and $N_x=12$ in other cases.
}
\label{fig:6}
\end{figure}

\subsection{Dephasing effects}\label{sec:dephasing}

Aside from the static impurities,
 dephasing is another significant impediment 
to robust performance. Then we look at the 
dephasing effects in the ZGNRs with MIS. The coherent length 
$L_{\phi}$ is a measure of coherence in experiments. Electrons can go from the 
left lead to the right lead directly or via the virtual leads in the presence of dephasing effects, resulting in coherent and incoherent currents, respectively. At a certain $dp$, the 
incoherent current grows as the length $N_x$ does as well. 
When the coherent and incoherent parts of the current are equal, 
$N_x$ is the coherent length $L_{\phi}$\cite{YXX}. Note that $L_{\phi}$ is an average value of the ZGNRs 
with MIS. The carbon-carbon distance in our model is not 
uniform along the $y$ axis. The real $L_{\phi}$ along 
the edge may be less than our numerical value.
$L_{\phi}$ changes with $dp$ for three distinct cases of 
EOMFs—PLLs, CLLs, and RLLs—are shown in Fig.~\ref{fig:6}(a). At a specific $dp$, 
the value of $L_{\phi}$ is the lowest for PLLs. That means dephasing 
effects are more sensitive to PMFs. There are two reasons for this. 
First, strain modifies the bond length between carbon 
atoms in the ZGNRs, and the dephasing effects caused by electron-phonon 
and electron-electron interactions are amplified\cite{enhance1,enhance2}. 
However, the cases of EOMFs ($\mathcal{B}_{r}\neq 0$) 
improve the value of $L_{\phi}$ in an obvious way. Because 
of the weakening of the hybridization 
demonstrated in Sec.~\ref{sec:disorders}, it can be concluded 
that the addition of RMFs considerably increases the sample's 
robustness against dephasing effects. 

In Fig.~\ref{fig:6}(b), 
we examine the conductance of the CLLs ($\mathcal{B}_{r}=50$~T) 
under various $dp$. 
In the weak and moderate dephasing regime ($dp<0.2$~eV), the 
conductance related to the lowest two CLLs are very robust due 
to the suppression of  hybridization. 
However, the ones with higher CLLs are not robust. This is 
due to the broadening of the higher CLLs brought on by the 
dephasing effects. In the higher CLLs, the spacing between 
adjacent energy levels gets smaller. Thus, the hybridization is 
again intensified. In the strong dephasing regime ($dp \ge 1 $~eV), the conductance 
exhibits a quasi-quantization with $G \lesssim 1$~eV. This 
outcome is in line with earlier researches~\cite{JPSJ1,chen1}. 
According to Fig.~\ref{fig:6}(c), the plateau of conductance 
gradually dissipates as the strain strength $\eta$ increases. 
The result is consistent with the performance of $L_{\phi}$ in 
Fig.~\ref{fig:6}(a). Finally, we consider the combination of 
Anderson disorder and dephasing effects in Fig.~\ref{fig:6}(d). 
The findings demonstrate that when disorder strength $W$ grows, 
conductance value marginally reduces under various $dp$. 

The results above show that tunable valley currents of the low CLLs 
are robust against dephasing effects when 
$\mathcal{B}_{r} \neq 0$. The edge states related to higher 
CLLs are not robust, though. Examining Hall conductance or 
developing novel solutions may be required as a next step to 
improve the performance of tuning valley currents against dephasing effects.

\begin{figure}[ht!]
\centering
\includegraphics[width=1.0\linewidth]{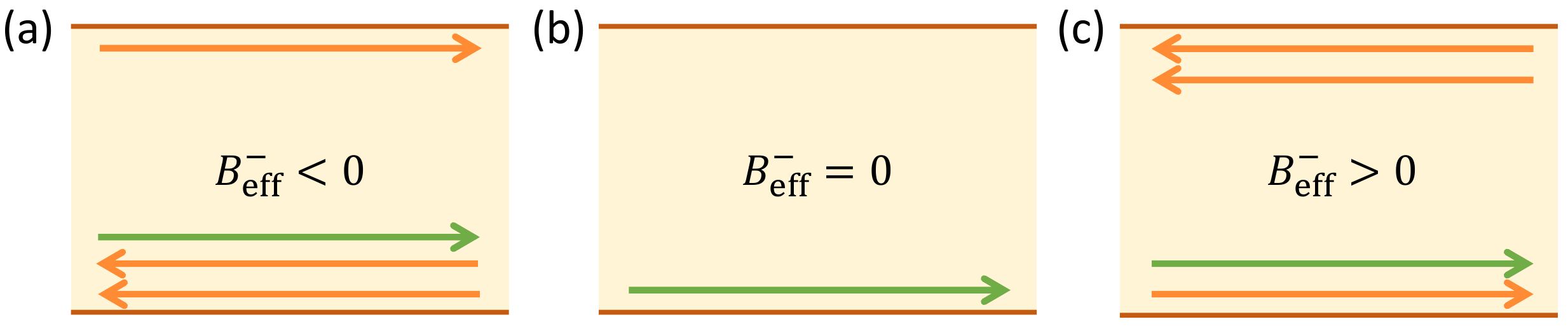}
\caption{The edge valley currents for (a) $\mathcal{B}^{-}_{\textrm{eff}}<0$, 
(b) $\mathcal{B}^{-}_{\textrm{eff}}=0$ and (c) 
$\mathcal{B}^{-}_{\textrm{eff}}>0$. The orange lines and green lines 
represent the currents in the $K'$ and $K$ valleys, respectively.
}
\label{fig:7}
\end{figure}

\subsection{Tunability of the valley-polarized currents}\label{sec:tunability}

The tunability of the valley-polarized currents in our system allows 
for the construction of new types of useful valleytronic devices. 
The band structures in Figs.~\ref{fig:2}(b)-(f) clearly demonstrate 
that the different EOMFs between the $K$ and $K'$ valleys induce the 
valley polarization. The valley currents depicted in 
Figs.~\ref{fig:7}(a)-(c) correspond to the states at Fermi energies 
$E_{1}$-$E_{3}$ in Figs.~\ref{fig:2}(b)-(d). The edge states are 
significantly out of balance between the $K$ and $K'$ valleys, 
which differ greatly from that of the well-known QHE, 
quantum spin Hall effects(QSHE), and QVHE. Therefore, our 
sample is a good platform for manufacturing valleytronic devices 
by tuning $E_F$, RMFs, or PMFs.

\begin{figure}[ht!]
\centering
\includegraphics[width=1.0\linewidth]{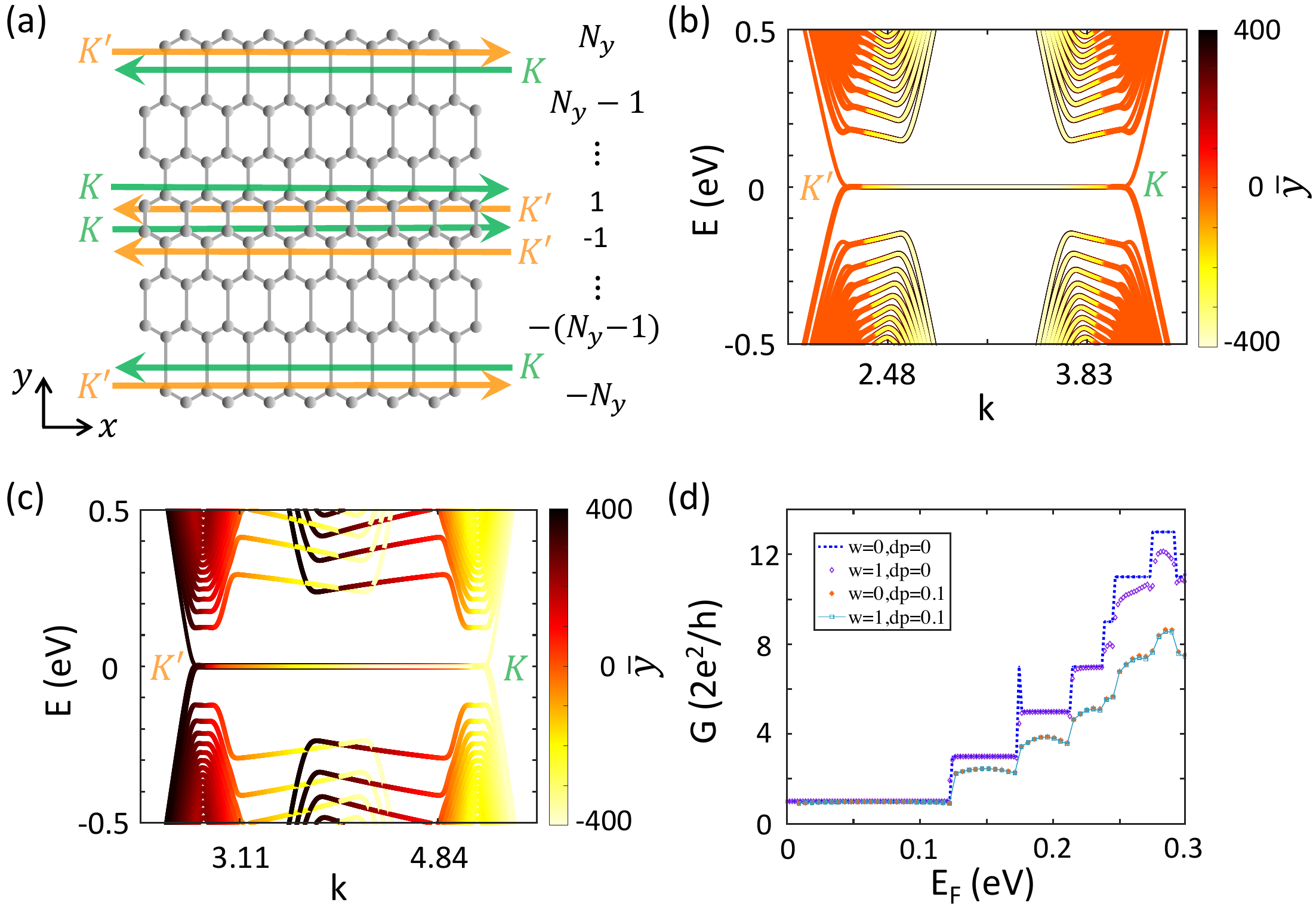}
\caption{(a) The schematic diagram of the ZGNRs with SS. (b) and (c) 
illustrate the band structures with $\mathcal{B}_{r}=0$ and $\mathcal{B}_{r}=50$~T, respectively. (d) The conductance $G$ vs the Fermi energy $E_F$ for various $W$ and $dp$ with $\mathcal{B}_{r}=50$ T. 
In all cases, we set $N_x=30$, $N_y=400$, and $\eta=0.5$.}
\label{fig:8}
\end{figure}

We may generalize our tunability of the valley-polarized  currents 
to the ZGNRs with symmetric strain (SS)~\cite{Bao}. As shown in 
Fig.~\ref{fig:8}(a), the ZGNRs with SS can be viewed as two reverse copies 
of the ZGNR with MIS, and snake states exist in the middle of the 
sample. Fig.~\ref{fig:8}(b) depicts the band structure of the ZGNR 
with SS, and previous work has demonstrated that without the RMFs, 
only the first plateau is robust against the Anderson disorders. If 
we apply a RMF $\mathcal{B}_{r}=50$~T, the CLLs are shown in 
Fig.~\ref{fig:8}(c), and as illustrated in Fig.~\ref{fig:8}(d), the higher 
plateaus related to the snake states are extremely robust against 
the Anderson disorders. In the presence of the dephasing effects, the 
plateaus of snake states are very robust for the first CLL and 
survive for the higher CLLs. Therefore, the results of the ZGNRs 
with SS are similar to that of the ZGNRs with MIS in 
Sec.~\ref{sec:disorders} and Sec.~\ref{sec:dephasing}. To sum up, 
snake states can survive for both Anderson disorders and dephasing 
effects which show excellent design potential for new quantum devices.

\section{Conclusions}\label{sec:conclusion}

In this work, we investigate the strained ZGNRs in the presence of 
the RMFs. The essential distinction between the RLLs and PLLs—which 
are produced by the RMFs and PMFs, respectively—is that the former 
are flat while the latter are dispersive. Because of their dispersive 
nature, the PLLs are susceptible to disorders because of the 
hybridization of their bulk and edge states. In order to incorporate 
the effects of the RMFs and PMFs, the concept of the EOMFs is introduced. 
Accordingly, we obtain the CLLs which combine the RLLs and PLLs. Then 
we examine the robust behaviors of the valley-polarized currents 
by calculating the conductance and discover that the RMF  
can improve the robust performances. Our transport 
calculations of the $K$ and $K'$ valleys demonstrate distinctive robust 
behaviors against the Anderson disorders and dephasing effect. 
The several mechanisms of the intervally and intravalley scattering make it clear how PLLs and RLLs behave differently on transport properties. Moreover, the valley polarization induced by the EOMFs $|\mathcal{B}^{\pm}_{\textrm{eff}}|$ reveals the distinct conductance performances that are related to the $K$ and $K'$ valleys, respectively. The behaviors of the conductance $G$ show that the RMF $\mathcal{B}_{r}$ is a valid tool for tuning the valley currents.
Furthermore, we investigate the tunability of the 
 valley-polarized currents in the ZGNRs with MIS and SS, respectively. 
We discover that our polarized edge states differ greatly from the 
well-known QHE, QSHE, and QVHE. Because of their efficient tunability, 
they are ideal for designing new sorts of valleytronics devices.

\begin{acknowledgments}

The authors would like to thank Qing-feng Sun, Shuai Zhao, Yanxia Xing, 
and Hua Jiang for helpful discussions. We are grateful to the National Natural Science Foundation of China(Grant No.12204053), 
the Natural Science Foundation of Shanghai (Grant No. 
21JC1402300) and Director's Fund of Key Laboratory of 
Polar Materials and Devices, Ministry of Education.

\end{acknowledgments}

\begin{appendix}

\section{The solutions of the CLLs}\label{sec:A1}

In this section, we use the similar method adopted in Ref.~\cite{Franz} 
to derive the expressions of CLLs. The Hamiltonian for the 
$K$ valley is shown in the Eq.~(\ref{eq:4}) in the main text:

\begin{align}\label{eq:A1}
\boldsymbol{d}^{+}\cdot\boldsymbol{\sigma}&=v_{F}\left(p_{x}+e\mathcal{B}_{\textrm{eff}}^{+}y\right)\sigma_{x} \nonumber \\
&-i\hbar v_{F}\left(r_{+}-s_{+}\frac{e\mathcal{B}_{p}y}{\hbar}\right)\partial_{y}\sigma_{y}.
\end{align}
We can shift $y$ to $y+\frac{\hbar r_{+}}{e\mathcal{B}_{p}s_{+}}$ 
and hermitize $-iy\partial_{y}$ to $-i\left(y\partial_{y}+\frac{1}{2}\right)$, 
then Eq.~(\ref{eq:A1}) becomes

\begin{align}\label{eq:A2}
\boldsymbol{d}^{+}\cdot\boldsymbol{\sigma}=\hbar v_{F}\left(\begin{array}{cc}
0 & h^{+}_{+}\\
h^{+}_{-} & 0
\end{array}\right),
\end{align}
where

\begin{align}\label{eq:A3}
h^{+}_{\pm}=
\left[k_{x}+\frac{e\mathcal{B}_{\textrm{eff}}^{+}}{\hbar}\left(y+\frac{\hbar r_{+}}{e\mathcal{B}_{p}s_{+}}\right)\right]\pm s_{+}\frac{e\mathcal{B}_{p}}{\hbar}\left(y\partial_{y}+\frac{1}{2}\right).
\end{align}
Thus the eigenvalue equations for the $K$ valley (Eq.~(\ref{eq:5}) in 
the main text) become 

\begin{align}\label{eq:A4}
&\left[k_{x}+\frac{e\mathcal{B}_{\textrm{eff}}^{+}}{\hbar}\left(y+\frac{\hbar r_{+}}{e\mathcal{B}_{p}s_{+}}\right)
+s_{+}\frac{e\mathcal{B}_{p}}{\hbar}\left(y\partial_{y}+\frac{1}{2}\right)\right]\psi_{B} \nonumber \\
&=\varepsilon\psi_{A}, \nonumber \\
&\left[k_{x}+\frac{e\mathcal{B}_{\textrm{eff}}^{+}}{\hbar}\left(y+\frac{\hbar r_{+}}{e\mathcal{B}_{p}s_{+}}\right)
-s_{+}\frac{e\mathcal{B}_{p}}{\hbar}\left(y\partial_{y}+\frac{1}{2}\right)\right]\psi_{A} \nonumber \\
&=\varepsilon\psi_{B}.
\end{align}
Eliminating $\psi_{A}$, we can obtain 

\begin{align}\label{eq:A5}
&\left[\left(\frac{e\mathcal{B}_{\textrm{eff}}^{+}}{\hbar}y\right)^{2}+\frac{e\mathcal{B}_{\textrm{eff}}^{+}}{\hbar s_{+}}\left(2k_{x}s_{+}+2r_{+}\frac{\mathcal{B}_{\textrm{eff}}^{+}}{\mathcal{B}_{p}}-\frac{e\mathcal{B}_{p}}{\hbar}s_{+}^{2}\right)y \right. \nonumber \\
&\left.+\frac{\Delta}{s_{+}^{2}}-\frac{e^{2}\mathcal{B}_{p}^{2}s_{+}^{2}}{4\hbar^{2}}\right]\psi_{B}-\frac{e^{2}\mathcal{B}_{p}^{2}}{\hbar^{2}}s_{+}^{2}\left(2y\psi'_{B}+y^{2}\psi''_{B}\right)=0,
\end{align}
where

\begin{align}\label{eq:A6}
\Delta=\left(k_{x}s_{+}+\frac{r_{+}\mathcal{B}_{\textrm{eff}}^{+}}{\mathcal{B}_{p}}\right)^{2}-s_{+}^{2}\varepsilon^{2}.
\end{align}
Two singularities of Eq.~\ref{eq:A5} are $0$ and $\infty$. 
Toward $0$, Eq.~\ref{eq:A5} can be asymptotically expressed 
as 

\begin{align}\label{eq:A7}
\left(\frac{\Delta}{s_{+}^{2}}-\frac{e^{2}\mathcal{B}_{p}^{2}s_{+}^{2}}{4\hbar^{2}}\right)\psi_{B}-\frac{e^{2}\mathcal{B}_{p}^{2}}{\hbar^{2}}s_{+}^{2}\left(2y\psi'_{B}+y^{2}\psi''_{B}\right)=0.
\end{align}
Eq.~\ref{eq:A7} has two independent solutions 
$y^{-\frac{1}{2}\mp\frac{\hbar\sqrt{\Delta}}{e\left|\mathcal{B}_{p}\right|s_{+}^{2}}}$, however, because we consider the asymptotic solution toward $0$, 
only $y^{-\frac{1}{2}+\frac{\hbar\sqrt{\Delta}}{e\left|\mathcal{B}_{p}\right|s_{+}^{2}}}$ 
is acceptable. Toward $\infty$, Eq.~\ref{eq:A5} is asymptotically 
expressed as

\begin{align}\label{eq:A8}
\left(\frac{e\mathcal{B}_{\textrm{eff}}^{+}}{\hbar}y\right)^{2}\psi{}_{B}-\left(\frac{e\mathcal{B}_{p}^{+}}{\hbar}s_{+}y\right)^{2}\psi''_{B}=0,
\end{align}
and the asymptotic solution can be written as $e^{-\frac{z}{2}}$, 
where

\begin{align}\label{eq:A9}
z=-\textrm{sgn}(\mathcal{B}_{p})\frac{2}{s_{+}}\left|\frac{\mathcal{B}_{\textrm{eff}}^{+}}{\mathcal{B}_{p}}\right|y.
\end{align}
Note that Eq.~\ref{eq:A9} differs from the asymptotic solution in 
Ref.~\cite{Franz} which contains solely PMFs. In conclusion, the 
general solution of Eq.~\ref{eq:A5} can be written as 

\begin{align}\label{eq:A10}
\psi_{B}(y)=e^{-\frac{z}{2}}y^{-\frac{1}{2}+\frac{\hbar\sqrt{\Delta}}{e\left|\mathcal{B}_{p}\right|s_{+}^{2}}}u(y).
\end{align}
Substitute Eq.~\ref{eq:A10} into Eq.~\ref{eq:A5}, and 
change the variable from $y$ to $z$, we can obtain the 
equation:

\begin{align}\label{eq:A11}
zu''(z)+\left(\xi-z\right)u'(z)-\alpha u(z)=0,
\end{align}
where 

\begin{align}\label{eq:A12}
\xi=&1+\frac{2\hbar\sqrt{\Delta}}{e\left|\mathcal{B}_{p}\right|s_{+}^{2}}, \nonumber \\
\alpha=&\frac{\sqrt{\Delta}-k_{x}s_{+}\textrm{sgn}(\mathcal{B}_{p}\mathcal{B}_{\textrm{eff}}^{+})-r_{+}|\frac{\mathcal{B}_{\textrm{eff}}^{+}}{\mathcal{B}_{p}}|}{e|\mathcal{B}_{p}|s_{+}^{2}/\hbar} \nonumber \\
&+\frac{\left(\mathcal{B}_{\textrm{eff}}^{+}+|\mathcal{B}_{\textrm{eff}}^{+}|\right)s_{+}^{2}}{2|\mathcal{B}_{\textrm{eff}}^{+}|s_{+}^{2}}.
\end{align}
Eq.~\ref{eq:A11} is a confluent hypergeometric equation, and 
$\alpha$ must be a nonpositive integer to guarantee that the solution 
is convergent. Thus $\alpha=-n$ leads to the CLLs

\begin{align}\label{eq:A13}
\varepsilon_{+}^{2}=ne\hbar v^2_{F}|\mathcal{B}_{\textrm{eff}}^{+}|\left(2k_{x}s_{+}\frac{\mathcal{B}_{p}}{\mathcal{B}_{\textrm{eff}}^{+}}+2r_{+}-n\frac{e\mathcal{B}_{p}^{2}}{\hbar|\mathcal{B}_{\textrm{eff}}^{+}|}s_{+}^{2}\right).
\end{align}

Because the third term is small, we can neglect it and obtain

\begin{align}\label{eq:A14}
\varepsilon_{+}^{2}=2ne\hbar v^2_{F}\left|\mathcal{B}_{\textrm{eff}}^{+}\right|\left(r_{+}+k_{x}s_{+}\frac{\mathcal{B}_{p}}{\mathcal{B}_{\textrm{eff}}^{+}}\right).
\end{align}
Similarly, the CLLs for the $K'$ valley are

\begin{align}\label{eq:A15}
\varepsilon_{-}^{2}=2ne\hbar v^2_{F}\left|\mathcal{B}_{\textrm{eff}}^{-}\right|\left(r_{-}+k_{x}s_{-}\frac{\mathcal{B}_{p}}{\mathcal{B}_{\textrm{eff}}^{-}}\right).
\end{align}
It should be pointed out that 
$\mathcal{B}_{\textrm{eff}}^{-}=\mathcal{B}_{r}-r_{+}\mathcal{B}_{p}$ 
can be zero because the directions of the RMF and PMF are opposite for 
the $K'$ valley. Actually, $\alpha$ for the $K'$ valley has the singular 
point $\mathcal{B}_{\textrm{eff}}^{-}=0$, as a result, our solutions for 
the CLLs are invalid when the EOMF $\mathcal{B}_{\textrm{eff}}^{-}=0$. 
Furthermore, $\varepsilon_{-}^{2}$ in Eq.~\ref{eq:A15} may be negative if 
the absolute value of $\mathcal{B}_{\textrm{eff}}^{-}$ is a small value. 
Therefore, our result for $\varepsilon_{-}$ is invalid in the vicinity 
of the singular point $\mathcal{B}_{\textrm{eff}}^{-}=0$. 

\section{Rashba spin-orbit coupling}\label{sec:A2}

\begin{figure}[ht!]
\centering
\includegraphics[width=1.0\linewidth]{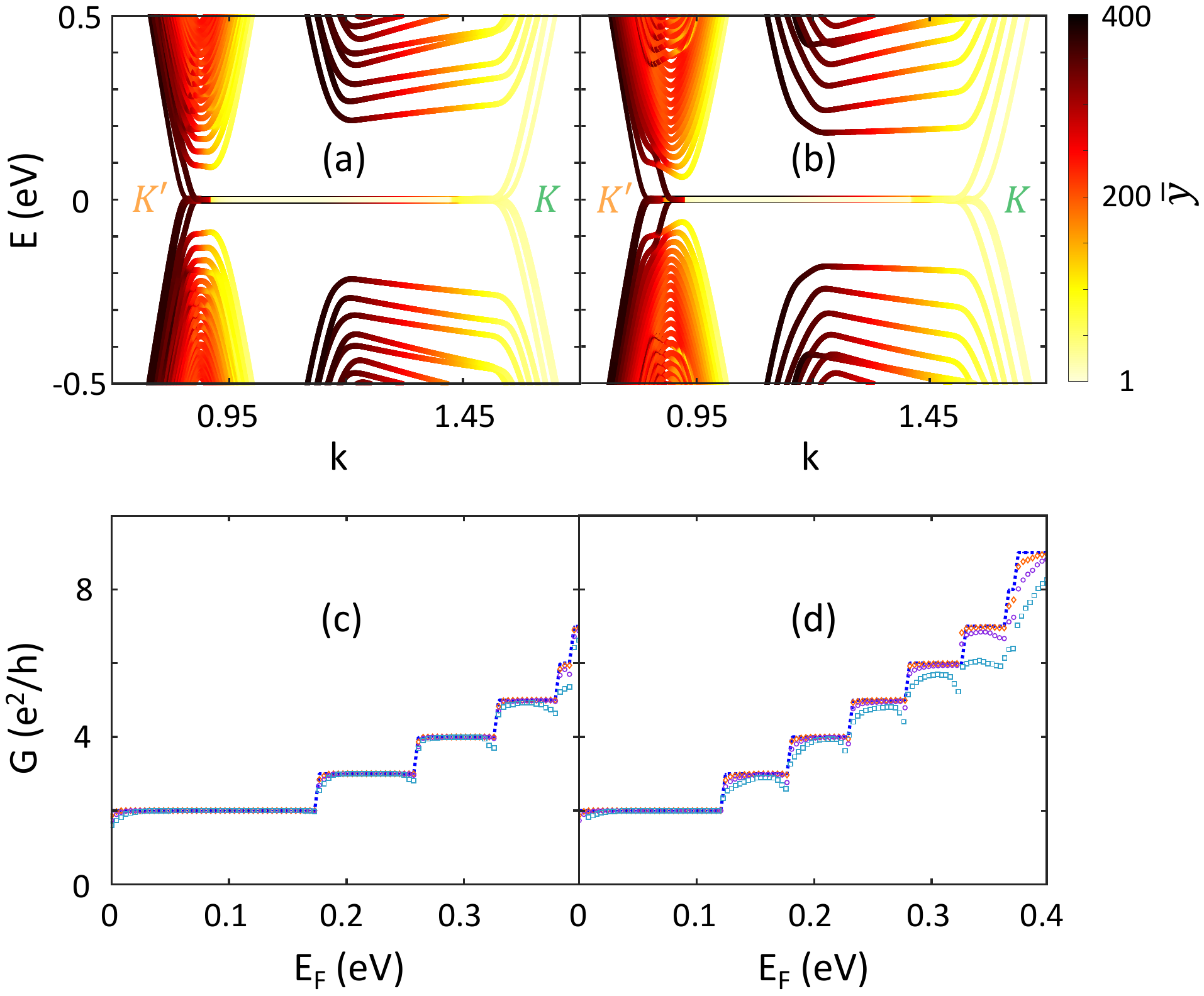}
\caption{(a) and (b) show the band structures with the RSOC 
strength $V_R=0.02t$ and $V_R=0.05t$, respectively. (c) and (d) are 
the conductance corresponding to (a) and (b). In all cases, we set 
$\mathcal{B}_{r}=50$~T and $\eta=0.5$. In all cases, $N_x=30$ and $N_y=200$.}
\label{fig:Rashba}
\end{figure}

We investigate the ZGNRs with MIS for various RSOC strengths 
$V_{R}$. It should be pointed out that since the RSOC breaks
the spin degeneracy, we take into account the spin degree of freedom 
in this part. The band structures and conductance 
for $V_{R}=0.02t$ and $V_{R}=0.05t$ are shown in Fig.~\ref{fig:Rashba},
demonstrating that the conductance are still quantized even if the 
RSOC causes the extended states. Additionally, odd plateaus 
appear as a result of the RSOC lifting the spin degeneration. 
Due to the existence of the RMFs, both the odd and even plateaus are 
robust against Anderson disorder. As a result, the RSOC does 
not invalidate the primary findings in the main paper.
Additionally, we ignore the Zeeman energy in our calculations 
because it is negligibly small and only splits the energy band, 
having no impact on the physics discussed in this paper.

\end{appendix}

\bibliographystyle{apsrev4-1}

\begin{thebibliography}{100}

\bibitem{grapheneRMP}
A. H. C. Neto, F. Guinea, N. M. R. Peres, K. S. Novoselov, and A. K. 
Geim, The electronic properties of graphene, 
\href{https://journals.aps.org/rmp/abstract/10.1103/RevModPhys.81.109}
{Rev. Mod. Phys. \textbf{81}, 109 (2009).}

\bibitem{Beenakker}
A. Rycerz, J. Tworzydło and C. W. J. Beenakker, Valley filter and 
valley valve in graphene, 
\href{https://www.nature.com/articles/nphys547}
{Nat. Phys. 3, 172 (2007).}

\bibitem{valley1}
S. P. Milovanovic\'{a} and F. M. Peeters, Strain controlled valley filtering 
in multi-terminal graphene structures,
\href{https://aip.scitation.org/doi/10.1063/1.4967977}
{Appl. Phys. Lett. \textbf{109}, 203108 (2016).}

\bibitem{valley2}
Z. Yu, F. Xu, and J. Wang, Valley Seebeck effect in gate tunable zigzag 
graphene nanoribbons,
\href{https://doi.org/10.1016/j.carbon.2015.12.033}
{Carbon \textbf{99}, 451 (2016).}

\bibitem{pLL1}
W.-Y. He and L. He, Coupled spin and pseudomagnetic field in graphene 
nanoribbons, 
\href{https://journals.aps.org/prb/abstract/10.1103/PhysRevB.88.085411}
{Phys. Rev. B \textbf{88}, 085411 (2013).}

\bibitem{pLL2}
F. Guinea, M. I. Katsnelson, and A. K. Geim, Energy gaps and a zero-field 
quantum Hall effect in graphene by strain engineering,
\href{https://www.nature.com/articles/nphys1420}
{Nat. Phys. \textbf{6}, 30 (2010).}

\bibitem{pLL3}
T. Low and F. Guinea, Strain-induced pseudomagnetic field for 
novel graphene electronics, 
\href{https://pubs.acs.org/doi/10.1021/nl1018063}
{Nano Lett. \textbf{10} 3551 (2010).}

\bibitem{pLL4}
M. A. H. Vozmediano, M. I. Katsnelson, and F.Guinea, 
Gauge fields in graphene, 
\href{https://www.sciencedirect.com/science/article/pii/S0370157310001729}
{Phys. Rep. \textbf{496}, 109 (2010).}

\bibitem{pLL5}
S. Zhu, J. A. Stroscio, and T. Li, Programmable extreme pseudomagnetic 
fields in graphene by a uniaxial stretch, 
\href{https://journals.aps.org/prl/abstract/10.1103/PhysRevLett.115.245501}
{Phys. Rev. Lett. \textbf{115}, 245501 (2015).}

\bibitem{pLL6}
D. Sabsovich, M. W. Bockrath, K. Shtengel, and E. Sela, Helical 
superconducting edge modes from pseudo-Landau levels in graphene, 
\href{https://journals.aps.org/prb/abstract/10.1103/PhysRevB.103.094513}
{Phys. Rev. B \textbf{103}, 094513 (2021).}

\bibitem{pLL7}
J. Sun, T. Liu, Y. Du, and H. Guo, 
Strain-induced pseudo-magnetic field in $\alpha-\mathcal{T}_{3}$ lattice, 
arXiv:2208.08136 (2022).

\bibitem{exp1}
N. Levy, S. A. Burke, K. L. Meaker, M. Panlasigui, A. Zettl, F. Guinea, 
A. H. C. Neto, and M. F. Crommie, Strain-induced pseudo–magnetic 
fields greater than 300 tesla in graphene nanobubbles, 
\href{https://www.science.org/lookup/doi/10.1126/science.1191700}
{Science \textbf{329}, 544 (2010).}


\bibitem{exp2}
Y. Liu, J. N. B. Rodrigues, Y. Z. Luo, L. Li, A. Carvalho, M. Yang, 
E. Laksono, J. Lu, Y. Bao, H. Xu, S. J. R. Tan, Z. Qiu, C. H. Sow, 
Y. P. Feng, A. H. C. Neto, S. Adam, J. Lu, and K. P. Loh, Tailoring 
sample-wide pseudo-magnetic fields on a graphene–black phosphorus 
heterostructure, 
\href{https://www.nature.com/articles/s41565-018-0178-z}
{Nat. Nanotechnology \textbf{13}, 828 (2018).}

\bibitem{exp3}
P. Nigge, A. C. Qu, \'{E}. Lantagne-Hurtubise, E. M$\mathrm{\mathring{a}}$rsell, 
S. Link, G. Tom, M. Zonno, M. Michiardi, M. Schneider, S. Zhdanovich, 
G. Levy, U. Starke, C. Guti\'{e}rrez, D. Bonn, S. A. Burke, M. Franz, and 
A. Damascelli, Room temperature strain-induced Landau levels in graphene 
on a wafer-scale platform, 
\href{https://www.science.org/doi/10.1126/sciadv.aaw5593}
{Sci. Adv. \textbf{5}, 5593 (2019).}

\bibitem{exp4}
S.-Y. Li, Y. Su, Y.-N. Ren, and L. He, Valley polarization and inversion 
in strained graphene via pseudo-Landau levels, valley splitting of real 
Landau levels, and confined states, 
\href{https://journals.aps.org/prl/abstract/10.1103/PhysRevLett.124.106802}
{Phys. Rev. Lett. \textbf{124}, 106802 (2020).}



\bibitem{strength1}
E. Cadelano, P. L. Palla, S. Giordano, and L. Colombo, Nonlinear elasticity 
of monolayer graphene, 
\href{https://journals.aps.org/prl/abstract/10.1103/PhysRevLett.102.235502}
{Phys. Rev. Lett. \textbf{102}, 235502 (2009).}

\bibitem{strength2}
C. Lee, X. Wei, J. W. Kysar, and J. Hone, Measurement of the elastic 
properties and intrinsic strength of monolayer graphene,
\href{https://www.science.org/doi/full/10.1126/science.1157996}
{Science \textbf{321}, 385 (2008).}

\bibitem{strength3}
F. Liu, P. Ming, and J. Li, Ab initio calculation of ideal strength 
and phonon instability of graphene under tension,
\href{https://journals.aps.org/prb/abstract/10.1103/PhysRevB.76.064120}
{Phys. Rev. B \textbf{76}, 064120 (2007).}

\bibitem{strength4}
I. Yu. Sahalianov, T. M. Radchenko, V. A. Tatarenko, G. Cuniberti, and 
Y. I. Prylutskyy, Straintronics in graphene: Extra large electronic 
band gap induced by tensile and shear strains, 
\href{https://aip.scitation.org/doi/10.1063/1.5095600}
{J. Appl. Phys. \textbf{126}, 054302 (2019).}



\bibitem{Franz}	
\'{E}tienne Lantagne-Hurtubise, X.-X. Zhang, and M. Franz, Dispersive 
Landau levels and valley currents in strained graphene nanoribbons, 
\href{https://journals.aps.org/prb/abstract/10.1103/PhysRevB.101.085423}
{Phys. Rev. B \textbf{101}, 085423 (2020).}


\bibitem{TB2}
V. M. Pereira, A. H. C. Neto, and N. M. R. Peres,Tight-binding approach to uniaxial strain in graphene
\href{https://journals.aps.org/prb/abstract/10.1103/PhysRevB.80.045401}
{Phys. Rev. B \textbf{80}, 045401(2009).}

\bibitem{HJiang}
B.-L. Wu, Q. Wei, Z.-Q. Zhang, and H. Jiang, Transport property of inhomogeneous strained graphene,
\href{http://cpb.iphy.ac.cn/EN/10.1088/1674-1056/abe3e3}
{Chin. Phys. B \textbf{30}, 030504(2021).}

\bibitem{Bao}
C.-y. Zuo, J. Qi, T.-l. Lu, Z.-q. Bao, and Y. Li, Reverse strain-induced 
snake states in graphene nanoribbons,
\href{https://journals.aps.org/prb/abstract/10.1103/PhysRevB.105.195420}
{Phys. Rev. B \textbf{105}, 195420 (2022).}

\bibitem{Buttiker}
M. B$\ddot{u}$ttiker, Role of quantum coherence in series resistors,
\href{https://doi.org/10.1103/PhysRevB.33.3020}
{Phys. Rev. B \textbf{33}, 3020 (1986).}


\bibitem{Landauer} 
R. Landauer, Electrical resistance of disordered one-dimensional lattices, 
\href{https://www.tandfonline.com/doi/abs/10.1080/14786437008238472}
{Phil. Mag., \textbf{21}, 863 (1970).}

\bibitem{Fisher1981} 
D. S. Fisher, and P. A. Lee, Relation between conductivity and transmission 
matrix,
\href{https://journals.aps.org/prb/abstract/10.1103/PhysRevB.23.6851}
{Phys. Rev. B \textbf{23}, 6851(R) (1981).}

\bibitem{Meir1992} 
Y. Meir and N. S. Wingreen, Landauer formula for the current through 
an interacting electron region, 
\href{https://journals.aps.org/prl/abstract/10.1103/PhysRevLett.68.2512}
{Phys. Rev. Lett. \textbf{68}, 2512 (1992).}
 
\bibitem{Jauho1994} 
A.-P. Jauho, N. S. Wingreen, and Y. Meir, Time-dependent transport in 
interacting and noninteracting resonant-tunneling systems,
\href{https://journals.aps.org/prb/abstract/10.1103/PhysRevB.50.5528}
{Phys. Rev. B, \textbf{50}, 5528 (1994).}

\bibitem{Sancho}
M. P. L. Sancho, J. M. L. Sancho and J. Rubio, Highly convergent 
schemes for the calculation of bulk and surface Green functions, 
\href{https://iopscience.iop.org/article/10.1088/0305-4608/15/4/009}
{J. Phys. F: Met. Phys. \textbf{15}, 851(1985).}

\bibitem{HJ}
H. Jiang, S. Cheng, Q.-f. Sun, and X. C. Xie, Topological Insulator: 
A New Quantized Spin Hall Resistance Robust to Dephasing,
\href{https://journals.aps.org/prl/abstract/10.1103/PhysRevLett.103.036803}
{Phys. Rev. Lett. \textbf{103}, 036803 (2009).}

\bibitem{YXX}
Y. Xing, Q.-f. Sun,  and J. Wang, Influence of dephasing on the quantum 
Hall effect and the spin Hall effect,
\href{https://journals.aps.org/prb/abstract/10.1103/PhysRevB.77.115346}
{Phys. Rev. B \textbf{77}, 115346 (2008)}.

\bibitem{Settnes}
M. Settnes, J. H. Garcia, and S. Roche, Valley-polarized quantum transport 
generated by gauge fields in graphene,
\href{https://iopscience.iop.org/article/10.1088/2053-1583/aa7cbd}
{2D Mater. \textbf{4}, 031006 (2017).}

\bibitem{enhance1}
M. B. Lundeberg and J. A. Folk, Rippled Graphene in an In-Plane Magnetic 
Field: Effects of a Random Vector Potential,
\href{https://journals.aps.org/prl/abstract/10.1103/PhysRevLett.105.146804}
{Phys. Rev. Lett. \textbf{105}, 146804 (2010)}.

\bibitem{enhance2}
R. Burgos, J, Warnes, Leandro R. F. Lima, and C. Lewenkopf, Effects 
of a random gauge field on the conductivity of graphene sheets with 
disordered ripples,
\href{https://journals.aps.org/prb/abstract/10.1103/PhysRevB.91.115403}
{Phys. Rev. B \textbf{91}, 115403 (2015)}.

\bibitem{JPSJ1}
Y. Shimomura and Y. Takane, Dephasing-Induced Stabilization of a 
Perfectly Conducting Channel in Disordered Graphene Nanoribbons 
with Zigzag Edges,
\href{https://journals.jps.jp/doi/10.7566/JPSJ.85.014704}
{J. Phys. Soc. Jpn. \textbf{85}, 014704 (2016).}

\bibitem{chen1}
J.-C. Chen, H. Zhang, S.-Q. Shen and Q.-f. Sun, Dephasing effect 
on transport of a graphene p–n junction in a quantum Hall regime, 
\href{https://iopscience.iop.org/article/10.1088/0953-8984/23/49/495301}
{J. Phys.: Condens. Matter \textbf{23} 495301 (2011).}

\end{thebibliography}

\end{document}